\documentclass[a4paper]{jpconf}
\usepackage{graphicx}

\usepackage{amssymb,amsopn}

\def\la{\lambda}

\def\ds{\stackrel{\star}{,}}
\def\x{\hat x}

\def\lb{\lbrack}
\def\rb{\rbrack}

\begin{document}


\title{Gauge theory on kappa-Minkowski revisited: the twist approach}

\author{Marija Dimitrijevi\' c$^1$ and Larisa Jonke$^2$}

\address{$^1$ University of Belgrade, Faculty of Physics,
Studentski trg 12, 11000 Beograd, Serbia}

\address{$^2$ Theoretical Physics Division, Rudjer Bo\v skovi\' c Institute,
Bijeni\v cka 54, 10000 Zagreb, Croatia}

\ead{dmarija@ipb.ac.rs, larisa@irb.hr}

\begin{abstract}
Kappa-Minkowski space-time is an example of noncommutative 
space-time with potentially interesting phenomenology. However, the construction of field theories on this space is plagued  with ambiguities. We propose to resolve certain ambiguities by clarifying the  geometrical picture of gauge transformations on the $\kappa$-Minkowski space-time in the twist approach.
 We construct the action for the noncommutative $U(1)$  gauge fields in a geometric way, as an integral of a maximal form. The effective action with  the first order corrections in the deformation parameter is obtained using the Seiberg-Witten map to relate noncommutative and commutative degrees of freedom.
\end{abstract}

\section{Introduction and overview}

It is generally believed that the picture of
space-time as a differentiable manifold should
break down at very short distances of the order
of the Planck length.  There are different
proposals for the modified space-time structure
which should provide a consistent framework
encompassing physics in this regime.  These
proposals include, among others, the dynamical
triangulation as a way of direct
geometrical construction of modified space-time, strings
and loops as non-local fundamental observables dynamically
generating space-time, and a deformation of algebra of functions on a manifold
as a way of introducing a 'noncommutative space-time'.

Introducing a  non-trivial algebra of coordinates in
order to modify the space-time structure can be seen as an attempt to  generalise the  concept of symmetries that should encompass physics on a quantum manifold. A possible way to describe physics on such a manifold is to construct (effective) field theory compatible with the algebra of coordinates in the framework of deformation quantization.
The main advantage of such effective models is that one can extract phenomenological
consequences of the space-time modification using  standard field-theoretical
tools. 

In this work our primary interest is to examine the compatibility of the local gauge principle with the deformation of the algebra of functions on a specific example of noncommutative
space-time, the $\kappa$-Minkowski space-time. The commutation relations of the
coordinates of this space-time are of the Lie-algebra type:
\begin{equation}
    \label{1.1}
[\hat{x}^0, \hat{x}^j] =\frac{i}{\kappa}\hat{x}^j,\; [\hat{x}^i, \hat{x}^j] = 0.
\end{equation}
 One of the interesting properties of this noncommutative
space-time is that there is a quantum group symmetry acting on it \cite{luk}. It is a dimensionful
deformation of the global Poincar\'e group, the $\kappa$-Poincar\'e group. The constant $\kappa$ has
dimension of energy and sets a deformation scale. The $\kappa$-Minkowski space also provides an arena for formulating new physical concepts. These include the generalisation of Special Relativity with one additional invariant scale known as Doubly Special Relativity \cite{DSR}, and the concept of relativity of locality \cite{RelLoc1}, a proposal in which space-time is observer-dependent projection from the invariant phase space. This makes the $\kappa$-Minkowski space-time  an example of noncommutative space-time
with potentially interesting phenomenology.

 We are interested in the construction of the 
field theory on this space as a  step towards extracting observable consequences of underlying noncommutative structure. In our previous
work \cite{miU1} we showed that  within the framework defined in \cite{kappaft,kappagft} one
can consistently describe a gauge theory on the $\kappa$-Minkowski space-time by explicit
construction of $U(1)$ gauge theory coupled to fermions. Although successful, the construction
revealed certain ambiguities which were fixed by  physical arguments and
intuition, rather then by the formalism itself.  With this motivation in mind, 
 we use the twist formalism in order to gain a better understanding of the gauge
theory on the $\kappa$-Minkowski space-time \cite{mi2}. Note, however, that in this formalism we cannot maintain the $\kappa$-Poincar\'e symmetry; the corresponding symmetry of the twisted $\kappa$-Minkowski space is the twisted $igl(1,3)$ symmetry.
One of the advantages of the twist formalism is the straightforward way to define a differential calculus. This enables us to write the  action  in a geometric way,
as an integral of a maximal form. Furthermore, the geometric description  we use clearly shows that one cannot decouple  translations and gauge symmetries, a generic feature for theories with underlying noncommutative structure. In our model this mixing of internal and space-time symmetries enters in the construction of the (Hodge) dual field strength.
We present two different ways of defining the dual field strength leading to the same expanded action in the first order in deformation parameter. 

\section{Noncommutative spaces from a twist}


 The main idea of the twist formalism is to first deform the symmetry of the
theory and then see the consequences this deformation has on the space-time
itself. There is a well defined way to deform the symmetry Hopf algebra. In his
paper \cite{Drinfeld} Drinfel'd introduced a notion of twist. The
twist ${\cal F}$ is an invertible operator which belongs to
$Ug \otimes Ug$, where $Ug$ is the
universal enveloping algebra of the symmetry Lie algebra $g$. The universal
enveloping algebra $Ug$ is a Hopf algebra
\begin{eqnarray}
[t^a, t^b] &=& if^{abc}t^c, \nonumber\\
\Delta(t^a) &=& t^a \otimes 1 + 1\otimes t^a, \nonumber\\
\varepsilon (t^a) &=& 0,\quad S(t^a) = -t^a .\label{Ug}
\end{eqnarray}
In the first line $t^a$ label the generators of the symmetry algebra $g$ and
the structure constants are labelled by $f^{abc}$. In the second line the coproduct of the generator $t^a$ is given. It
encodes the Leibniz rule and specifies how the symmetry transformation acts on products
of fields/representations. In the last line, the counit and the
antipode are given. The properties which the twist $\cal{F}$ has
to satisfy are:
\begin{enumerate}
\item the cocycle condition
\begin{equation}
\label{propF11}
({\cal F}\otimes 1)(\Delta\otimes id){\cal F}=(1\otimes
{\cal F})(id\otimes \Delta){\cal F}, \label{Twcond1}
\end{equation}
\item normalization
\begin{equation}
(id\otimes \epsilon){\cal F} = (\epsilon\otimes id){\cal F}=1\otimes 1,
\label{Twcond2}
\end{equation}
\item perturbative expansion
\begin{equation}
{\cal F} = 1\otimes 1 + {\cal O}(\lambda), \label{Twcond3}
\end{equation}
\end{enumerate}
where $\lambda$ is a small deformation parameter. The last property is
not necessary. It provides an expansion around the undeformed case in the limit
 $\lambda\to 0$. We shall frequently use the notation (sum over $\alpha=1,2,...\infty $
is understood)
\begin{equation}
\label{Fff}
{\cal F} = {\rm f}^\alpha \otimes{\rm f}_\alpha, \quad
{\cal F}^{-1}=\bar{\rm f}^\alpha \otimes\bar{\rm f}_\alpha ,
\end{equation}
where, for each value of $\alpha$, $\bar{\rm f}^\alpha$ and
$\bar{\rm f}_\alpha$ are two distinct elements of $Ug$ (and similarly
${\rm f}^\alpha$ and ${\rm f}_\alpha$ are in $Ug$).
We also introduce the universal $\cal{R}$-matrix
\begin{equation}
{\cal R} = {\cal F}_{21}{\cal F}^{-1} ,\label{defUR}
\end{equation}
where by definition ${\cal F}_{21}={\rm f}_{\alpha}\otimes
{\rm f}^\alpha$.
In the sequel we use the notation
\begin{equation}
{\cal R} = R^\alpha\otimes R_\alpha ,\quad
{\cal R}^{-1} = \bar{R}^\alpha\otimes\bar{R}_\alpha.
\end{equation}


The twist acts on the symmetry Hopf algebra and gives the twisted
symmetry Hopf algebra
\begin{eqnarray}
[t^a, t^b] &=& if^{abc}t^c,\nonumber\\
\Delta_{\cal F}(t^a) &=& {\cal F}\Delta (t^a){\cal F}^{-1} \nonumber\\
\varepsilon(t^a) &=& 0, \quad
S_{\cal F}(t^a) = {\rm f}^\alpha S({\rm f}_\alpha)S(t^a)S(\bar{\rm f}^\beta)
\bar{\rm f}_\beta. \label{TwistedUg}
\end{eqnarray}
We see that the algebra remains the same, while in general the
comultiplication changes. This leads to the deformed Leibniz rule for the
symmetry transformations when acting on product of fields.

We can now use the twist to
deform the commutative geometry on space-time (vector fields, one-forms,
exterior algebra of forms, tensor algebra).
The guiding principle is the observation that every time we have
a bilinear map
$$\mu\,: X\times Y\rightarrow Z ,$$
where $X,Y,Z$ are vector spaces
and when there is an action of the Lie algebra $g$
(and therefore of ${\cal F}^{-1}$) on $X$ and $Y$
we can combine the map $\mu$ with the action of the twist. In this way
we obtain the deformed map $\mu_\star$:
\begin{equation}
\mu_\star = \mu {\cal F}^{-1}. \label{mustar}
\end{equation}
The cocycle condition (\ref{propF11}) implies that if $\mu$ is an
associative product then also $\mu_\star$ is an associative product.

Let us analyze this deformation in more detail. For convenience we 
now consider one particular class of twists, the Abelian twists
\begin{equation}
{\cal F} = e^{-\frac{i}{2}\theta^{ab}X_a\otimes X_b} .\label{AbTwist} 
\end{equation}
Here $\theta^{ab}$ is a constant antisymmetric matrix
and $X_a=X_a^\mu\partial_\mu$ are commuting vector fields.
The algebra of vector fields on the space-time $M$ we label with $\Xi$
and the universal enveloping algebra of this algebra with $U\Xi$. 
Then ${\cal F}$ belongs to $U\Xi \otimes U\Xi$. In the view of 
(\ref{Ug})-(\ref{Twcond3}), the symmetry algebra is the algebra of
diffeomorphisms generated by vector fields $\xi=\xi^\mu\partial_\mu\in \Xi$. 
Note that depending on the choice of vector fields $X_a$ one can also
consider a subalgebra of the diffeomorphism algebra such as Poincar\' e
or conformal algebra.

Applying the inverse of the twist (\ref{AbTwist}) to the usual 
point-wise multiplication of functions on the space-time $M$, 
$\mu(f\otimes g)=f\cdot g$, we obtain the $\star$-product of
functions
\begin{eqnarray}
f\star g &=& \mu {\cal F}^{-1}(f\otimes g)\nonumber\\
&=& \bar{\rm f}^\alpha(f) \bar{\rm f}_\alpha(g) \nonumber\\
&=& \bar{R}^\alpha(g)\star\bar{R}_\alpha(f). \label{FunctionsStar}
\end{eqnarray}
We see that the $R$-matrix encodes the
noncommutativity of the $\star$-product.
The action of the twist ($\bar{{\rm f}}^\alpha$ and $\bar{{\rm f}}_\alpha$) on the functions $f$ and $g$ is via the Lie derivative.

The product between functions and one-forms is given by following
the general prescription
\begin{equation}
h\star\omega = \bar{\rm f}^\alpha(h)\bar{\rm f}_\alpha(\omega)
\end{equation}
with an arbitrary one-form $\omega$. The action of $\bar{{\rm f}}_\alpha$ on
forms is given via the Lie derivative.
Functions can be multiplied from the left or from the right,
\begin{equation}
h\star\omega = \bar{\rm{f}}^\alpha(h)\bar{\rm{f}}_\alpha(\omega)
={\bar{R}^\alpha}(\omega)\star \bar{R}_{\alpha}(h).
\label{FunFormStar}
\end{equation}

Exterior forms form an algebra with the wedge product
$\wedge :\,\Omega^{\mbox{\boldmath $\cdot$}}\times
\Omega^{\mbox{\boldmath $\cdot$}}\rightarrow \Omega^{\mbox{\boldmath $\cdot$}}$.
We $\star$-deform the wedge product on two arbitrary forms $\omega$
and $\omega'$ into the $\star$-wedge product,
\begin{equation}
\omega\wedge_\star\omega' = \bar{\rm f}^\alpha(\omega)
\wedge \bar{\rm f}_\alpha(\omega').\label{WedgeStar}
\end{equation}
We denote by $\Omega^{\mbox{\boldmath $\cdot$}}_\star$
the linear space of forms equipped with the $\star$-wedge product
$\wedge_\star$.

As in the commutative case exterior forms are totally
$\star$-antisymmetric (contravariant) tensor-fields.
For example, two-form
$\omega\wedge_\star\omega'$ is the $\star$-antisymmetric combination
\begin{eqnarray}
\omega\wedge_\star\omega' &=& \bar{\rm f}^\alpha(\omega)
\wedge \bar{\rm f}_\alpha(\omega') \nonumber\\
&=&\omega\otimes_\star\omega'
-\bar{R}^\alpha(\omega')\otimes_\star \bar{R}_{\alpha}(\omega),
\label{Star1Forms}\\
&=& -\bar{R}^\alpha(\omega')\wedge_\star \bar{R}_{\alpha}(\omega) ,
\nonumber
\end{eqnarray}
with the $\star$-tensor product defined as
\begin{equation}
T_1\otimes_\star T_2 = \bar{\rm f}^\alpha(T_1)
\otimes \bar{\rm f}_\alpha(T_2). \label{TensPrStar}
\end{equation}

The usual exterior derivative
${\rm d}: {\cal A}_x \rightarrow \Omega$, as it commutes with the Lie derivative, satisfies the Leibniz rule 
${\rm d} (f\star g) = {\rm d} f\star g + f\star {\rm d}g $ and is
therefore also the $\star$-exterior derivative. One can rewrite the
usual exterior derivative of a function using the $\star$-product as
\begin{eqnarray}
{\rm d}f &=& (\partial_\mu f){\rm d}x^\mu \nonumber\\
&=& (\partial^\star_\mu f)\star {\rm d}x^\mu, \label{ExtDeriv}
\end{eqnarray}
where the new derivatives $\partial_\mu^\star$ are defined by this
equation.

The usual integral is cyclic under the $\star$-exterior products of
forms, that is up to boundary terms we have
\begin{equation}
\int \omega_1 \wedge_\star \omega_2 = (-1)^{d_1\cdot d_2}
\int \omega_2 \wedge_\star \omega_1, \label{IntCycl}
\end{equation}
where $d=deg(\omega)$, $d_1+d_2 = m$ and $m$ is the dimension of the
space-time $M$. This property holds for the Abelian twist 
(\ref{AbTwist}). More generally, one can show \cite{PL1} that this 
property holds for any twist that satisfies the condition 
$S(\bar{\rm f}^\alpha)\bar{\rm f}_\alpha =1$,
with the antipode $S$.

\section{Kappa-Minkowski via twist}

Algebraically, the four-dimensional $\kappa$-Minkowski space-time can be
introduced as a  quotient
of the algebra freely generated by coordinates $\hat{x}^\mu$ divided by
the ideal generated by the following commutation relations:
\begin{equation}
\label{2.1}
\lb \x^\mu , \x^\nu \rb = i C^{\mu\nu}_\rho \x^\rho,
\quad \mu,\nu,\rho=0,\dots,3.
\end{equation}
Defining
\begin{equation}
\label{2.2}
 C^{\mu\nu}_\rho=a(\delta^\mu _0 \delta^\nu_\rho
-\delta^\nu_0\delta^\mu_\rho)
\end{equation}
the commutation relations (\ref{2.1}) can be rewritten as
\begin{equation}
[\hat{x}^0, \hat{x}^j] = ia\hat{x}^j, \quad
[\hat{x}^i, \hat{x}^j] = 0. \label{Intro2Kom}
\end{equation}
The metric of the $\kappa$-Minkowski space-time is
$\eta^{\mu\nu}=diag(1,-1,-1,-1)$.
The deformation parameter $a$ is related to the
frequently used parameter $\kappa$ as $a=1/\kappa$.
Latin indices denote the space dimensions, zero the time
dimension and the Greek indices refer to all  dimensions.

 The choice of twist is not unique and it depends on the
properties that we want to obtain/preserve.  We choose the following twist
\begin{eqnarray}
{\cal F} &=& e^{-\frac{i}{2}\theta^{ab}X_a\otimes X_b}\nonumber\\
&=& e^{-\frac{ia}{2}
(\partial_0\otimes x^j\partial_j-x^j\partial_j\otimes \partial_0)},
\label{KappaTwist}
\end{eqnarray}
with two commuting vector fields $X_1=\partial_0$ and $X_2=x^j\partial_j$
and
\begin{equation}
\theta^{ab} =\left( {\begin{array}{cc}
 0 &  a \\ -a &  0
\end{array} } \right)
. \nonumber
\end{equation}
Our choice is motivated by the fact that the twist (\ref{KappaTwist}) leads (see below) to the hermitean $\star$-product and the cyclic integral, the properties crucial for the construction of an action.  
One can check that this twist fulfils the conditions (\ref{Twcond1}), (\ref{Twcond2}) and
(\ref{Twcond3})
with the small deformation parameter $\lambda=a$. 
Deformed symmetry concerned, note that $X_2$ is not in
the universal enveloping algebra of the Poincar\' e algebra. Therefore we
have to enlarge the Poincar\' e algebra $iso(1,3)$ to the inhomogeneous general
linear algebra $igl(1,3)$ and twist this algebra instead of 
$iso(1,3)$. The generators (given in the representation on the
space of functions/fields) and the commutation relations
of $igl(1,3)$ are
\begin{eqnarray}
&& M_{\mu\nu} = x_\mu\partial_\nu , \quad P_\mu =\partial_\mu,\nonumber\\
&& \lbrack P_\mu, P_\nu \rbrack = 0, \quad
\lbrack M_{\mu\nu}, P_\rho\rbrack = \eta_{\mu\rho}P_\nu,\nonumber\\
&& \lbrack M_{\mu\nu}, M_{\rho\sigma}\rbrack = \eta_{\nu\rho}M_{\mu\sigma}
- \eta_{\mu\sigma}M_{\rho\nu} . \label{igl(1n)}
\end{eqnarray}

Let us discuss the consequences of the twist (\ref{KappaTwist}).
 The action of the twist (\ref{KappaTwist}) on the $igl(1,3)$
algebra follows from (\ref{TwistedUg}) and it has been analysed in detail
in \cite{Pachol}. Here we just summarise the most important results.
The algebra (\ref{igl(1n)}) remains the same.
On the other hand, since $X_2 = x^j\partial_j$ does not commute with
the generators $\partial_\mu$ and $M_{\mu\nu}$ the comultiplication
and the antipode change. Here we just give the result for the twisted
comultiplication, the other results can be found in \cite{Pachol},
\begin{eqnarray}
\Delta P_0 &=& P_0 \otimes 1 + 1\otimes P_0, \nonumber\\
\Delta P_j &=& P_j \otimes e^{-\frac{i}{2}aP_0}
+ e^{\frac{i}{2}aP_0}\otimes P_j, \nonumber\\
\Delta M_{ij} &=& M_{ij} \otimes 1 + 1\otimes M_{ij}, \nonumber\\
\Delta M_{0j} &=& M_{0j} \otimes e^{-\frac{i}{2}aP_0}
+ e^{\frac{i}{2}aP_0}\otimes M_{0j} -\frac{i}{2}aP_j\otimes \mathrm{D}
+ \frac{i}{2}a\mathrm{D}\otimes P_j, \nonumber\\
\Delta M_{j0} &=& M_{j0} \otimes e^{-\frac{i}{2}aP_0}
+ e^{\frac{i}{2}aP_0}\otimes M_{j0} ,\nonumber\\
\Delta M_{00} &=& M_{00} \otimes 1 + 1\otimes M_{00}
-\frac{i}{2}aP_0\otimes \mathrm{D}
+ \frac{i}{2}a\mathrm{D}\otimes P_0. \label{TwistedCopr}
\end{eqnarray}
We introduced the notation $\mathrm{D}=x^j\partial_j$. Note that
$\kappa$-Poincar\' e symmetry
found in \cite{luk} will not be a symmetry of our twisted
$\kappa$-Minkowski space. The corresponding symmetry of the twisted
$\kappa$-Minkowski space is the twisted $igl(1,3)$ symmetry.
The twisted symmetry does not have 
the usual dynamical significance and there is no Noether procedure associated with it. We view this symmetry as a way of bookkeeping, a prescription that allow us to consistently apply deformation in the theory.

The inverse of the twist (\ref{KappaTwist}) defines the $\star$-product
between functions/fields on the $\kappa$-Minkowski space-time
\begin{eqnarray}
f\star g &=& \mu_\star \{ f\otimes g \} \nonumber\\
&=& \mu \{ {\cal F}^{-1}\, f\otimes g\} \label{StarDef}\\
&=& \mu \{ e^{\frac{ia}{2}
(\partial_0\otimes x^j\partial_j-x^j\partial_j\otimes \partial_0)}
f\otimes g\}
\nonumber\\
&=& f\cdot g + \frac{ia}{2} x^j\big( (\partial_0 f) \partial_j g
-(\partial_j f) \partial_0 g\big) + {\cal O}(a^2) \nonumber\\
&=& f\cdot g + \frac{i}{2}C^{\rho\sigma}_\lambda x^\lambda
(\partial_\rho f)\cdot (\partial_\sigma g) + {\cal O}(a^2)
, \label{StarPrExp}
\end{eqnarray}
with $C^{\rho\sigma}_\lambda$ given in (\ref{2.2}).
This product is associative, noncommutative and hermitean
\begin{equation}
\overline{f\star g} = \bar{g} \star \bar{f}. \nonumber
\end{equation}
The usual complex conjugation we label with ``bar''. In the zeroth order
(\ref{StarPrExp}) reduces to the usual point-wise multiplication. Of course, we obtain
\begin{equation}
[x^0 \ds x^j] = x^0\star x^j -
x^j \star x^0 = ia x^j,\quad [x^i \ds x^j] =0 .\label{xStarComm}
\end{equation}

One of the advantages of the twist formalism is the straightforward way
to define a differential calculus. Namely, as said in the previous section,
we just adopt the undeformed differential calculus with the following
properties
\begin{eqnarray}
{\rm d} (f\star g) &=& {\rm d}f\star g + f\star {\rm d}g,\nonumber\\
{\rm d}^2 &=& 0,\nonumber\\
{\rm d} f &=& (\partial_\mu f) {\rm d}x^\mu = (\partial^\star_\mu f)
\star {\rm d}x^\mu. \label{Differential}
\end{eqnarray}
The basis one-forms are ${\rm d}x^\mu$. Knowing that the action of a
vector field on a form is given via Lie derivative one can show that
\begin{equation}
X_1 ({\rm d}x^\mu) =0, \quad
X_2 ({\rm d}x^\mu) = \delta ^\mu_j {\rm d}x^j.\label{LieDerdx}
\end{equation}
Using these relations one obtains that the basis one-forms anticommute but do not
$\star$-commute with functions. Instead they fulfil
\begin{eqnarray}
{\rm d}x^\mu \wedge_\star {\rm d}x^\nu &=& {\rm d}x^\mu \wedge {\rm d}x^\nu
= - {\rm d}x^\nu \wedge {\rm d}x^\mu =  -{\rm d}x^\nu \wedge_\star
{\rm d}x^\mu,
\nonumber\\
f\star {\rm d}x^0 &=& {\rm d}x^0 \star f,\quad
f\star {\rm d}x^j = {\rm d}x^j \star e^{ia\partial_0}f. \label{fstardx}
\end{eqnarray}
Arbitrary one-forms $\omega_1= \omega_{1\mu}\star {\rm d}x^\mu$ and
$\omega_2= \omega_{2\mu}\star {\rm d}x^\mu$ do not anticommute
\begin{equation}
\omega_1\wedge_\star \omega_2 =
- \bar{R}^\alpha(\omega_2)\wedge_\star \bar{R}_{\alpha}(\omega_1),
\label{anticom1forme}
\end{equation}
where the inverse of the ${\cal R}$ matrix  is given by
\begin{equation}
{\cal R}^{-1} = {\cal F}^2 = e^{-ia(\partial_0\otimes x^j\partial_j -
x^j\partial_j\otimes \partial_0)}. \label{Rbar}
\end{equation}
The $\star$-derivatives follow from (\ref{Differential}) and are given by
\begin{eqnarray}
&&\partial^\star_0 = \partial_0, \quad \partial^\star_j
= e^{-\frac{i}{2}a\partial_0}\partial_j,\nonumber\\
&&\partial^\star_0 (f\star g) = (\partial^\star_0 f)\star g +
f\star (\partial^\star_0 g), \nonumber\\
&& \partial^\star_j (f\star g) =
(\partial^\star_j f)\star e^{-ia\partial_0}g +
f\star (\partial^\star_j g). \label{ParcLeibniz}
\end{eqnarray}
The usual integral of a maximal form is cyclic
\begin{equation}
\int \omega_1 \wedge_\star \omega_2 = (-1)^{d_1\cdot d_2}
\int \omega_2 \wedge_\star \omega_1, \label{kappaIntCycl}
\end{equation}
with $d_1+d_2 = 4$. Since basis one-forms anticommute the volume
form remains undeformed
\begin{equation}
{\rm d}^{4}_\star x := {\rm d}x^0\wedge_\star{\rm d}x^1\wedge_\star
\dots {\rm d}x^{3} = {\rm d}x^0\wedge{\rm d}x^1\wedge\dots
{\rm d}x^{3} = {\rm d}^{4} x .\label{VolForm}
\end{equation}

\section{$U(1)$ gauge theory}

In this section we formulate pure noncommutative $U(1)$ gauge theory on the twisted $\kappa$-Minkowski space-time. The coupling to matter was analysed in \cite{mi2}. 
We start by introducing the  noncommutative connection $$A = A_\mu \star {\rm d}x^\mu,$$
written in the coordinate basis.  The transformation law
 of the noncommutative connection  is
given by
\begin{equation}
\delta_\alpha^\star A = {\rm d}\Lambda_\alpha + i[\Lambda_\alpha \ds A],
\label{ATr}
\end{equation}
or in the components
\begin{eqnarray}
\delta_\alpha^\star A_0 &=& \partial_0 \Lambda_\alpha
+ i [\Lambda_\alpha \ds A_0], \label{AnTr}\\
\delta^\star_\alpha A_j &=& \partial_j^\star \Lambda_\alpha
+ i\Lambda_\alpha\star A_j
- iA_j\star e^{-ia\partial_0}\Lambda_\alpha. \label{AjTr}
\end{eqnarray}
The field-strength
tensor is a two-form given by
\begin{equation}
F = \frac{1}{2} F_{\mu\nu} \star {\rm d}x^\mu \wedge_\star {\rm d}x^\nu
= {\rm d}A - i A\wedge_\star A, \label{F}
\end{equation}
or in components
\begin{eqnarray}
F_{0j} &=& \partial_0^\star A_j - \partial_j^\star A_0
-iA_0\star A_j + iA_j\star e^{-ia\partial_0}A_0
,\label{Fnj}\\
F_{ij} &=& \partial_i^\star A_j - \partial_j^\star A_i
-iA_i\star e^{-ia\partial_0}A_j + iA_j\star e^{-ia\partial_0}A_i .
\label{Fij}
\end{eqnarray}
One can check that (\ref{F}) transforms covariantly,
\begin{equation}
\delta_\alpha^\star F = i [\Lambda_\alpha \ds F]. \label{FTr}
\end{equation}

As a next step we would like to construct the action. In the undeformed, commutative gauge theory\footnote{In the following, the undeformed, commutative fields will be denoted by superscript $(0)$, e.g., $F^{(0)}_{\alpha\beta}=\partial_\alpha A^{(0)}_\beta-\partial_\beta A^{(0)}_\alpha$, is the usual $U(1)$ field strength.} one writes the action for the gauge
field using the Hodge dual of the field-strength tensor $*F^{(0)}$
\begin{eqnarray}
&&S^{(0)} = -\frac{1}{2}\int F^{(0)}\wedge (*F^{(0)}), \nonumber\\
&&*F^{(0)} = \frac{1}{2}\epsilon_{\mu\nu\alpha\beta}F^{{(0)}\alpha\beta}
{\rm d}x^\mu\wedge {\rm d}x^\nu.\nonumber
\end{eqnarray}
The indices on $F^{{(0)}\alpha\beta}$ are raised with the flat metric
$\eta_{\mu\nu}$ and
\begin{equation}
\delta_\alpha (*F^{(0)}) = i [\alpha, *F^{(0)}] = 0 \nonumber
\end{equation}
since we work with $U(1)$ gauge theory.

We try to generalise this to the $\kappa$-Minkowski space-time. We write the
noncommutative gauge field action as
\begin{equation}
S =  c_1\int F\wedge_\star (*F), \label{NCSg}
\end{equation}
where $*F$ is the noncommutative Hodge dual field strength. In order to have an
action invariant under the noncommutative gauge transformations
(\ref{ATr}) the dual field strength has to transform covariantly
\begin{equation}
\delta^\star_\alpha (*F) = i [\Lambda_\alpha \ds *F] .\label{*FTr}
\end{equation}
The obvious guess for the noncommutative Hodge dual
\begin{equation}
*F = \frac{1}{2}\epsilon_{\mu\nu\alpha\beta} F^{\alpha\beta}\star
{\rm d}x^\mu\wedge_\star {\rm d}x^\nu \label{*Fobvious}
\end{equation}
does not work since it does not transform covariantly
\begin{equation}
\delta^\star_\alpha (*F)
= {1\over 2}\epsilon_{\mu\nu\alpha\beta} (\delta^\star_\alpha
F^{\alpha\beta})\star {\rm d}x^{\mu}\wedge_\star {\rm d}x^{\nu}
\neq i [\Lambda_\alpha \ds *F] .\nonumber
\end{equation}
Therefore we have to try something else. We
assume that $*F$ has the form
\begin{equation}
*F := {1\over 2}\epsilon_{\mu\nu\alpha\beta}
X^{\alpha\beta}\star {\rm d}x^{\mu}\wedge_\star {\rm d}x^{\nu} ,
\label{*FX}
\end{equation}
where $X^{\alpha\beta}$ are unknown components that should  be determined
form the condition (\ref{*FTr}).
Unfortunately, we were unable to find consistent Ansatz for  $X^{\alpha\beta}(A_{\mu})$ in the closed form. Up to the first order in the  deformation parameter we find: 
\begin{eqnarray}
X^{0j} &=& {F}^{0j}-aA_0\star{F}^{0j},\nonumber \\
X^{jk} &=& {F}^{jk}+aA_0\star{F}^{jk}. \label{Xcomp}
\end{eqnarray}
Inserting this into (\ref{*FX}) gives dual field strength that does transform covariantly under the gauge transformations.

Going back to the action (\ref{NCSg}) and writing it more explicitly
we obtain
\begin{equation}
S = -\frac{1}{4}\int \Big\{ 2F_{0j}\star e^{-ia\partial_0}X^{0j}
+ F_{ij}\star e^{-2ia\partial_0}X^{ij}\Big\}\star {\rm d}^4 x .
\label{SgComp}
\end{equation}
where the components of $F$ and $X$ are given in (\ref{Fnj}), (\ref{Fij}) and
(\ref{Xcomp}). The terms $e^{-ia\partial_0}X^{0j}$ and
$e^{-2ia\partial_0}X^{ij}$ come from $\star$-commuting the basis
one-forms with the components $X^{\mu\nu}$. The constant $c_1$ is fixed in
such a way as to give the good commutative limit of the action
(\ref{SgComp}).

\section{Seiberg-Witten map}

Next, we would like to extract from the action (\ref{SgComp}) the first order corrections in the deformation parameter. To this end we use the Seiberg-Witten (SW) map \cite{SW,mssw,EnvAlg}. The idea behind the SW map is that the noncommutative gauge transformation is induced by the commutative one, $\delta_{\alpha}\to \delta_{\alpha}^\star$. This means that we can express the noncomutative fields and gauge parameter as functions of the commutative ones, e.g., $\Lambda_\alpha = \Lambda_\alpha (A_\mu^{(0)})$, $A_\mu=A_\mu(A_\mu^{(0)})$. 
In this way the number of degrees
of freedom in the noncommutative theory reduces to the number of degrees
of freedom of the corresponding commutative theory.
Demanding that the algebra of gauge transformations,
defined in (\ref{ATr}), closes gives the consistency condition:
\begin{equation}
(\delta^\star_\alpha \delta^\star_\beta -\delta^\star_\beta
\delta^\star_\alpha )A =  \delta^\star_{-i[\alpha,\beta]}A
.\label{swalg}
\end{equation}
This condition is taken as an  equation giving the expression for noncommutative gauge parameter $\Lambda_\alpha$ in terms of the commutatative gauge parameter $\alpha$ and the commutative gauge field $A_\mu^{(0)}$. In general, the solution can not be obtained in the closed form. Expanding the gauge parameter $\Lambda_\alpha$ in the orders of the deformation parameter   enables one to solve the equation order by order. In this paper we are interested in the first order correction, therefore we  construct the SW map only up to first order. 
Writing $\Lambda_\alpha$ as 
\begin{equation}
\Lambda_\alpha = \alpha + \Lambda_\alpha^1+ \dots + \Lambda_\alpha^k
+\dots ,\nonumber
\end{equation}
assuming $\Lambda_\alpha^k = \Lambda_\alpha^k (A_\mu^{(0)})$,  and
expanding the $\star$-product in the equation (\ref{swalg}), we obtain the inhomogeneous equation
for $\Lambda_\alpha^1$:
\begin{eqnarray}
&& \delta_\alpha\Lambda^1_\beta -\delta_\beta\Lambda^1_\alpha =
-C^{\rho\sigma}_\lambda x^\lambda
(\partial_\rho \alpha) (\partial_\sigma\beta)
.\label{Lam1}
\end{eqnarray}
Note that $\delta_\alpha\Lambda^1_\beta \neq 0$ since $\Lambda^1_\beta$
is a function of the commutative gauge parameter $\beta$ and the commutative
gauge field $A_\mu^{(0)}$ and $\delta_\alpha A_\mu^{(0)}
= \partial_\mu \alpha \neq 0$. The solution of equation (\ref{Lam1}) is given by
\begin{equation}
\Lambda^1_{\alpha} = -\frac{1}{2}C^{\rho\sigma}_\lambda x^\lambda
A^{(0)}_\rho \partial_\sigma\alpha. \nonumber
\end{equation}
This solution is not unique, one can always add a solution of the
homogeneous equation to it. This is the freedom in the SW map. In the
case of $U(1)$ gauge group the only homogeneous term is of the form
\begin{equation}
\Lambda_\alpha^{\rm{hom}} = b_1C^{\rho\sigma}_\lambda x^\lambda
F_{\rho\sigma}^{(0)}\alpha .\nonumber
\end{equation}
However, this term does not lead to a solvable equation for the noncommutative
gauge field and therefore we shall not consider it. The noncommutative gauge parameter up
to first order in the deformation parameter reads
\begin{equation}
\Lambda_{\alpha} = \alpha -\frac{1}{2}C^{\mu\nu}_\lambda x^\lambda
A^{(0)}_\mu \partial_\nu\alpha. \label{SWLambda}
\end{equation}

In order to find the SW map for gauge field $A_\mu(A_\mu^{(0)})$, we assume that $A_\mu = A_\mu^0 + A_\mu^1 + \dots$, insert this expansion into (\ref{AnTr}) and (\ref{AjTr}), and  expand the $\star$-product. Thus obtained equation for  $A_\mu(A_\mu^{(0)})$ we solve up to the first order we obtain:
\begin{eqnarray}
A_\mu &=& A^{(0)}_\mu - \frac{a}{2}\delta^j_{\mu}\Big(i\partial_0 A^{(0)}_j
+ A^{(0)}_0A^{(0)}_j\Big)+ {1\over 2} C^{\rho\sigma}_\lambda x^\lambda
\Big(  F^{(0)}_{\rho \mu}A^{(0)}_\sigma
- A^{(0)}_\rho \partial _\sigma A^{(0)}_\mu\Big) \nonumber\\
&& + d_1 C^{\rho\sigma}_\la x^\la \partial_\rho F^{(0)}_{\sigma\mu}
+ d_2 a F_{\mu 0}^{(0)}.\label{SWAmu}
\end{eqnarray}
The terms with the real undetermined coefficients $d_1$ and $d_2$ are the solutions of the
homogeneous equation and represent the freedom of the SW map.
Note that the connection one-form $A$ is real, but the components
$A_\mu$ are not necessarily real due to the $\star$-product in
$A = A_\mu\star {\rm d}x^\mu$.

Inserting the solution (\ref{SWAmu}) into (\ref{Fnj}) and (\ref{Fij})
results in the SW map for the field strength tensor:
\begin{eqnarray}
F_{0j} &=& F_{0j}^{(0)} -\frac{ia}{2}\partial_0F_{0 j}^{(0)} -aA_0^{(0)}F_{0j}^{(0)}
+ C^{\rho\sigma}_\lambda x^\lambda\Big( F^{(0)}_{\rho 0}F_{\sigma j}^{(0)}
-A^{(0)}_\rho\partial _\sigma F_{0j}^{(0)} \Big) \nonumber\\
&&  + a(d_1-d_2) \partial_0 F_{0j}^{(0)}
,\label{SWFnj}\\
F_{ij} &=& F_{ij}^{(0)} -ia\partial_0 F_{ij}^{(0)} - 2aA_0^{(0)}F_{ij}^{(0)}
+ C^{\rho\sigma}_\lambda x^\lambda \Big( F^{(0)}_{\rho i}F_{\sigma j}^{(0)}
- A^{(0)}_\rho\partial _\sigma F_{ij}^{(0)} \Big)
\nonumber\\
&& + a(d_1-d_2)\partial_0F_{ij}^{(0)} . \label{SWFij}
\end{eqnarray}

\section{Equations of motion and expanded action}

Having defined the action and the SW map for the gauge fields  we are ready to calculate the equations of motion for the fields. We expand the action (\ref{SgComp}) in the first order in the deformation parameter using the SW map and expanding the $\star$-product. This gives an effective action for the undeformed gauge fields with the first order corrections coming from the deformation we introduced:
\begin{eqnarray}
&& S^{\rm eff} = -\frac{1}{4}\int {\rm d}^4 x\Big \{ F^{(0)}_{\mu\nu}F^{{(0)}\mu\nu}
-\frac{1}{2}C^{\rho\sigma}_\lambda x^\lambda F^{{(0)}\mu\nu}F^{(0)}_{\mu\nu}
F^{(0)}_{\rho\sigma}+  2C^{\rho\sigma}_\lambda x^\lambda F^{{(0)}\mu\nu}
F^{(0)}_{\mu\rho}F^{(0)}_{\nu\sigma}\Big \}. \label{SgExp}
\end{eqnarray}
Note that there are no ambiguous terms in the expanded action coming from the freedom in the SW map; all such terms turned out to be total derivative terms and therefore they dropped out from the expanded action. The equation of motion for  the gauge field is:
\begin{eqnarray}
&&\partial_\mu F^{{(0)}\alpha\mu} =- \frac{a}{4}\delta^\alpha_0
F^{{(0)}\mu\nu}F^{(0)}_{\mu\nu} + 2a F^{{(0)}\alpha \mu}F^{(0)}_{0\mu}
\!+\! C^{\rho\sigma}_\lambda x^\lambda
\Big( F_{\rho}^{{(0)}\mu}\partial_\mu F_{\sigma}^{{(0)}\alpha}
\!+\! F^{(0)}_{\mu\sigma}\partial_{\rho}F^{{(0)}\mu\alpha} \Big)
 .\label{ExpEOMAmu}
\end{eqnarray}
We see that there is no modification of the dispersion relation for the free photon field $A_\mu^{(0)}$ in the first order of the deformation parameter. And this result does agree with our previous findings, see analysis in \cite{Bolokhov}.
However, the $x$-dependent terms in our expanded action clearly 
demand better understanding, possibly in terms of geometric degrees of 
freedom. Furthermore, one needs to understand the renormalization properties 
of the theory before making any predictions. Based on  
the results obtained in field theory on the canonically deformed 
space-time one does expect  additional terms in the action  which 
render theory renormalizable \cite{NCRenorm}. Finally, 
the second order corrections in the deformation parameter might turn 
out to be essential for deforming of the dispersion relations.

\section{$U(1)$ gauge theory - take two}

We introduced the noncommutative  $U(1)$ gauge theory in Section 4, 
where we have chosen to work in 
the coordinate basis. The principal advantage of doing the 
explicit calculations in the coordinate basis is that one works with the 
flat metric. However, the fact that the basis one-forms ${\rm d}x_\mu$ 
do not commute with  functions (\ref{fstardx}) presented  a 
(technical) obstruction in the construction of the gauge-covariant 
expression for the Hodge dual field strength.    
In this section we discuss the construction of the noncommutative 
$U(1)$ gauge theory using so-called natural/nice/central basis 
\cite{Alex} for the explicit calculations. The main advantage of 
working in this basis is that the basis one-forms do commute with 
functions.

We change from the coordinate basis
\begin{eqnarray}
&&x^\mu=(t=x^0, x, y, z), \quad {\rm d}x^\mu
= ({\rm d}t, {\rm d}x, {\rm d}y, {\rm d}z),\quad \partial_\mu
= (\partial_t, \partial_x, \partial_y, \partial_z) \nonumber
\end{eqnarray} 
to the nice basis which is given by
\begin{eqnarray}
&& x^a = (t, r, \theta, \varphi), \quad \theta^a=({\rm d}t,
\frac{{\rm d}r}{r}, {\rm d}\theta, {\rm d}\varphi), \quad
e_a=(\partial_t, r\partial_r, \partial_\theta, \partial_\varphi) .
\label{nicebasis}
\end{eqnarray}
We rewrite the  twist (\ref{StarPrExp}) in the new basis as 
\begin{eqnarray}
{\cal F} = e^{-\frac{i}{2}\theta^{ab}X_a\otimes X_b}= e^{-\frac{ia}{2}
(\partial_t\otimes r\partial_r-r\partial_r\otimes \partial_t)} \label{NiceTwist}
\end{eqnarray}
with $X_1=\partial_t=e_0$ and $X_2=x^j\partial_j=r\partial_r=e_1$.
Consequently, the $\star$-product between the functions is now given as:
\begin{eqnarray}
f\star g &=& \mu {\cal F}^{-1}(f\otimes g)\nonumber\\
&=& f\cdot g +\frac{ia}{2}\big( (e_0f)(e_1g) - (e_1f)(e_0g)\big) 
+{\cal O}(a^2)  \nonumber\\
&=& f\cdot g +\frac{ia}{2}\big( (\partial_t f)(r\partial_r g) 
- (r\partial_r f)(\partial_t g)\big) 
+{\cal O}(a^2) . \label{NiceStar}
\end{eqnarray}
As we already mention, the $\star$-product between the functions and the basis one-forms $\theta^a$ is trivial
\begin{equation}
f\star \theta^a= \theta ^a\star f = f\cdot \theta^a. \label{NiceFunForms}
\end{equation}
This is a consequence of the fact that the Lie-derivatives along the vector fields $X_1$  and $X_2$ commute with the basis one-forms; ${\cal L}_{e_0} \theta^a = {\cal L}_{e_1} \theta^a =0$.
However, the new basis is not flat, and the metric is given by 
$g_{ab} = diag(1, -r^2, -r^2, -r^2\sin^2\theta)$. The twist 
(\ref{NiceTwist}) is semi-Killing since the metric does not depend on 
$t$. This in particular means that the $\star$-inverse of the metric 
tensor $g_{ab}$ is the 
same as the usual inverse $g_{ab}\star g^{ac} = g_{ab} g^{ac} 
= \delta_a^c$. The volume element is
\begin{equation}
d^4x= \sqrt{-g} \epsilon_{abcd}\theta^a\wedge\theta^b\wedge\theta^c
\wedge\theta^d = r^2\sin\theta{\rm d}t{\rm d}r{\rm d}\theta{\rm d}
\varphi .\label{NiceVol}
\end{equation}    

 The gauge field or the connection  
$A = A_a \theta^a$ is  one-form and under the infinitesimal noncommutative  gauge 
transformations it transforms as
\begin{eqnarray}
\delta^\star_\alpha A &=& {\rm d}\Lambda_\alpha + i[\Lambda_\alpha \ds A]\nonumber\\
\delta^\star_\alpha A_a &=& e_a\Lambda_\alpha + i[\Lambda_\alpha \ds A_a].\label{NiceDeltaA}
\end{eqnarray}
The last line follows from (\ref{NiceFunForms}). The field-strength 
tensor is defined as usual
and it transforms covariantly under the noncommutative  gauge transformations
\begin{eqnarray}
F &=& {\rm d}A  - i A\wedge_\star A, \quad F_{ab} = e_a A_b - e_b A_a -i [A_a\ds A_b] \nonumber\\
\delta^\star_\alpha F &=& i[\Lambda_\alpha \ds F], \quad \delta^\star_\alpha F_{ab} 
= i[\Lambda_\alpha \ds F_{ab}].\label{NiceF}
\end{eqnarray}  
The Hodge dual field-strength tensor we define generalizing the usual expression for the Hodge dual in curved space given by
\begin{equation}
*F^{(0)} = \frac{1}{2}\epsilon_{abcd}\sqrt{-g}g^{ae}g^{bf}F^{(0)}_{ef}\theta^c\wedge\theta^d.
\end{equation} 
Since we want that $*F$ transforms covariantly under the  noncommutative  gauge 
transformations we have to covariantize the metric. More precisely, 
we have to covariantize the whole expression $\sqrt{-g}g^{ae}g^{bf}$. 
Let us define
\begin{equation}
*F = \frac{1}{2}\epsilon_{abcd}G^{aebf}\star F_{ef}\star 
\theta^c\wedge_\star\theta^d. \label{Nice*F}
\end{equation} 
Here $G^{aebf}$ is the quantity that under noncommutative gauge transformations transforms covariantly 
\begin{equation}
\delta^\star_\alpha G^{aebf} = i[\Lambda_\alpha \ds G^{aebf}] 
,\label{NiceTrLawG}
\end{equation}
and in the limit $a\to 0$ it reduces to $\sqrt{-g}g^{ae}g^{bf}$.
Using the Seiberg-Witten map for $\Lambda_\alpha$ rewritten in the new basis
\begin{equation}
\Lambda_\alpha = \alpha - \frac{a}{2}(A^{(0)}_0(e_1 \alpha) 
- A^{(0)}_1(e_0 \alpha)), \label{NiceSWLambda} 
\end{equation}
 and expanding the 
$\star$-products in equation
(\ref{NiceTrLawG}) the solution for $G^{aebf}$ up to first order in $a$ 
follows
\begin{equation}
 G^{aebf} = \sqrt{-g}g^{ae}g^{bf} - aA^{(0)}_0e_1(\sqrt{-g}g^{ae}g^{bf})
. \label{NiceSWG}
\end{equation}
For completeness, let us write the Seiberg-Witten map solutions for $A$ and $F$ in the nice basis: 
\begin{eqnarray}
A_a &=& A_a^{(0)} + \frac{a}{2}\Big( A_1^{(0)}F^{(0)}_{0a} -A_0^{(0)}F^{(0)}_{1a}
+ A_1^{(0)}(e_0A_a^{(0)}) - A_0^{(0)}(e_1A_a^{(0)}) \Big), \label{NiceSWA}\\
F_{ab} &=& F^{(0)}_{ab} +a \Big( F^{(0)}_{0a}F^{(0)}_{1b} - F^{(0)}_{1a}F^{(0)}_{0b}
-A^{(0)}_0(e_1 F^{(0)}_{ab}) +A^{(0)}_1(e_0 F^{(0)}_{ab}) \Big)
. \label{NiceSWF}
\end{eqnarray}

Having all these results at hand, we write the gauge 
invariant action for pure $U(1)$ gauge theory on $\kappa$-Minkowski as:
\begin{equation}
S_g = -\frac{1}{2}\int (*F)\wedge_\star F, \label{NiceAction} 
\end{equation}
with $F$ and $*F$ given by (\ref{NiceSWF}) and (\ref{Nice*F}) 
respectively. Expanding this action up to first order in $a$ we obtain
\begin{equation}
S = -\frac{1}{4}\int {\rm d}^4 x\Big \{ F^{(0)}_{ab}F^{{(0)}ab}
+ aF^{{(0)}ab}(4F^{(0)}_{0a}F^{(0)}_{1b} - F^{(0)}_{01}F^{(0)}_{ab}) 
 \Big \} , \label{NiceActionExp} 
\end{equation}
with ${\rm d}^4 x = r^2\sin\theta{\rm d}t{\rm d}r{\rm d}\theta{\rm d}
\varphi$. We used that 
\begin{eqnarray}
&& \int (\partial_r f){\rm d}t{\rm d}r{\rm d}\theta{\rm d}
\varphi =  \int (r\partial_r f){\rm d}t\frac{{\rm d}r}{r}
{\rm d}\theta{\rm d}\varphi = {\mbox{ surface term}} =0,\nonumber \\
&& \theta^a\wedge\theta^b\wedge\theta^c
\wedge\theta^d = \epsilon^{abcd}{\rm d}t\wedge\frac{{\rm d}r}{r}
\wedge{\rm d}\theta\wedge {\rm d}\varphi .
\end{eqnarray}
Note that the action (\ref{NiceActionExp}) has the same form as the 
 first order expanded
action for the noncommutative  gauge field in the case of $\theta$-constant 
deformation. This is the consequence of the particular choice of basis 
in which the twist looks like the Moyal-Weyl twist. Note that
the change of coordinates (\ref{nicebasis}) cannot be done globally since it is not well defined in the origin $x^\mu =0$.

Finally we would like to compare the result (\ref{NiceActionExp}) with
the expanded action we obtained in the coordinate basis. The change
from one basis to the other is done via the matrix $L$, for example
\begin{equation}
{\rm d}x^\mu = L^\mu_{\,\, a} \theta^a. \nonumber 
\end{equation}
We find
\begin{equation}
L^\mu_{\,\, a} = \left( \begin{array}{cccc}
1&0&0&0\\
0&x& \frac{xz}{\sqrt{x^2+y^2}} & -y\\
0&y& \frac{yz}{\sqrt{x^2+y^2}} & x\\
0&z&-\sqrt{x^2 +y^2}&0
\end{array} \right) .\label{NiceTransfMatrix}
\end{equation}
Then
\begin{eqnarray}
F^{{(0)}ab}F^{(0)}_{0a}F^{(0)}_{1b} &=& x^jF^{{(0)}\mu\nu}F^{(0)}_{0\mu}F^{(0)}_{j\nu}, 
\nonumber\\ 
F^{{(0)}ab}F^{(0)}_{01}F^{(0)}_{ab} &=& x^jF^{{(0)}\mu\nu}F^{(0)}_{0j}F^{(0)}_{\mu\nu}
\nonumber .
\end{eqnarray}
Replacing this into (\ref{NiceActionExp}) we conclude that the 
action is the same in both basis, as expected. Note that there is 
a subtlety here. We used the SW map to render the  expression 
$\sqrt{-g}g^{ae}g^{bf}$ covariant, and in solving the condition 
(\ref{NiceTrLawG}) we omitted the  possible ambiguous terms 
which solve the homogeneous part of the equation. The reason 
for this is that in this approach we do not have control over 
these ambiguities. One should have constructed the SW map for 
metric $g^{bf}$ with the ambiguous terms included and then calculate 
the SW map for the expression $\sqrt{-g}g^{ae}g^{bf}$. An easier route might be expressing the metric in terms of vielbeins and constructing the complete SW map for them, but this we leave for future work.

\section{Conclusion and outlook}

In this paper we used the twist formalism to gain a better
understanding of the gauge theory on $\kappa$-Minkowski space-time and to
resolve certain ambiguities we encounter in our previous analysis
\cite{miU1}. The twist formalism provided us with a naturally
defined differential calculus. As a consequence,
we obtained uniquely defined derivatives, thus solving one ambiguity.
Next, in the twist approach the integral has the trace property,
and there is no need to introduce an additional measure
function in the integral. This also means that the limit
of vanishing deformation parameter $a$ reproduces the undeformed
case without the need for additional field redefinitions.
One puzzling feature of the gauge field on $\kappa$-Minkowski
disclosed within the formalism introduced in \cite{kappagft},
was that a gauge field is given in terms of the higher order differential
operator.
This produced "torsion-like"
terms in the field strength which were simply
omitted in the constructed action. In the twist
approach however, the commutation rule of the  basis
one-forms with functions reproduces the effect of
"higher order differential operator" gauge field without producing
unwanted terms in the action. 

We have shown that the twisting of symmetries, as a
way of deforming the algebra of coordinates, is compatible
with the local gauge principle.  The obstruction we have encountered in the construction of the Hodge-dual field-strength tensor is a manifestation of the fact that  the
introduction of a noncommutative geometrical structure prevents decoupling of translation and gauge symmetries. We proposed two different possibilities for construction of the dual field strength, both leading to the same effective action in the first order of the deformation parameter. 

Although the mixing of space-time and internal symmetries appeared as a problem in our construction, this is in fact one of the most intriguing property of models based on non-trivial algebras of coordinates.  
One possible way to understand this mixing was offered in the 
framework of Yang-Mills type matrix models \cite{Steinacker}. 
There it was shown that $U(1)$ part of general $U(N)$ gauge group 
can be interpreted as induced gravity coupling to the rest ($SU(N)$) 
gauge degrees of freedom. It would be interesting to see if such an 
interpretation is possible in our framework, by constructing models with larger gauge groups.

\ack
We would like to thank  Anna Pachol for fruitful discussion. We gratefully acknowledge the support of ESF within the framework of the Research Networking
Programme on "Quantum Geometry and Quantum Gravity" and the support of
SEENET-MTP Network through the ICTP-SEENET-MTP grant PRJ-09.  L. J.
acknowledges the support of the Ministry of
Science, Education and Sport of the Republic
of Croatia under the contract 098-0982930-2861. The work of M.D. is supported by
Project No.171031 of the Serbian Ministry of Education and Science.

\end{document}